\documentclass[twocolumn,showpacs,amsmath,amssymb,prl,superscriptaddress,floatfix]{revtex4-1}
\usepackage{graphicx}
\usepackage{bm}
\usepackage{bbold}
\usepackage{amsmath,amsfonts,amssymb}

\begin{document}

\title{Dynamical phase transitions as a resource for quantum enhanced metrology}

\author{Katarzyna Macieszczak}
\affiliation{School of Mathematical Sciences, University of
Nottingham, Nottingham, NG7 2RD, UK}
\affiliation{School of Physics and Astronomy, University of
Nottingham, Nottingham, NG7 2RD, UK}
\author{M\u{a}d\u{a}lin Gu\c{t}\u{a}}
\affiliation{School of Mathematical Sciences, University of
Nottingham, Nottingham, NG7 2RD, UK}
\author{Igor Lesanovsky}
\author{Juan P. Garrahan}
\affiliation{School of Physics and Astronomy, University of
Nottingham, Nottingham, NG7 2RD, UK}


\date{\today}

\begin{abstract}
We consider the general problem of estimating an unknown control parameter of an open quantum system.  We establish a direct relation between the evolution of both system and environment and the precision with which the parameter can be estimated.  We show that when the open quantum system undergoes a first-order dynamical phase transition the quantum Fisher information (QFI), which gives the upper bound on the achievable precision of any measurement of the system and environment, becomes quadratic in observation time (cf.\ ``Heisenberg scaling'').  In fact, the QFI is identical to the variance of the dynamical observable that characterises the phases that coexist at the transition, and enhanced scaling is a consequence of the divergence of the variance of this observable at the transition point.  This identification allows to establish the finite time scaling of the QFI. Near the transition the QFI is quadratic in time for times shorter than the correlation time of the dynamics.  In the regime of enhanced scaling the optimal measurement whose precision is given by the QFI involves measuring both system and output.  As a particular realisation of these ideas, we describe a theoretical scheme for quantum enhanced phase estimation using the photons being emitted from a quantum system near the coexistence of dynamical phases with distinct photon emission rates.
\end{abstract}

\maketitle

\noindent
{\em Introduction.}
The estimation of unknown parameters is a crucial task for quantum technology applications such as state tomography~\cite{Blatt}, system identification~\cite{SI}, and quantum metrology~\cite{metrologyreview,GI,spectroscopy}.  Enhancement in precision can be achieved by using highly correlated/entangled quantum states which encode the unknown parameter, like the Greenberger-Horne-Zeilinger (GHZ) state $|{\rm GHZ}\rangle = |0\rangle^{\otimes N} + |1\rangle^{\otimes N}$ constructed out of $N$ qubits.  With such a state as a resource an unknown parameter $g$ can be encoded as $|{\rm GHZ}_{g}\rangle = |0\rangle^{\otimes N} + e^{-i N g} |1\rangle^{\otimes N}$.  Since the phase effectively encoded in the state is $N g$, the estimation error on $g$ scales as $N^{-2}$ (referred to as Heisenberg scaling \cite{Heisenberg}) instead of the standard $N^{-1}$ scaling for a non-correlated state $(|0\rangle + e^{-ig}|1\rangle )^{\otimes N}$.

\begin{figure*}[ht!]
\begin{center}
\includegraphics[width=1.9\columnwidth]{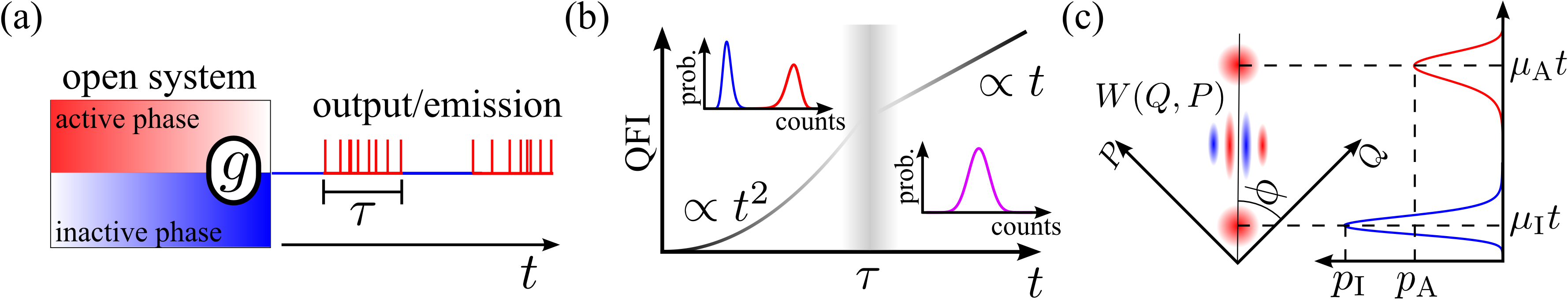}
\caption{(a) Open quantum system with a dynamics that features two dynamical phases of different activity and depends on the unknown parameter $g$. Near a first-order DPT the output (e.g. photons) shows strong intermittency where the temporal length of active/inactive periods is approximately given by the correlation time $\tau$. (b) In the vicinity of a DPT the QFI of the combined system-output state scales quadratically for observation times $t\ll \tau$. In the example of photon counting this regime features a bimodal count distribution, i.e. the two phases can be resolved. For $t\gg \tau$ this is no longer the case and the distribution becomes unimodal. Consequently, the QFI acquires a linear scaling with $t$. (c) Wigner distribution $W(Q,P)$ of the state (\ref{eq:coh}) after being projected on an appropriate system state, e.g. $| {\rm I} \rangle+| {\rm A} \rangle$. The two peaks are located at radii that correspond to the count rates $\mu_{\rm I,A}$ of the inactive/active phase. The count distribution is not sensitive to the parameter $\phi$ and hence counting is not an appropriate measurement for phase estimation. Still, this state features an enhanced QFI with respect to changes in the parameter $\phi$ due to the highly oscillatory fringe pattern [with period $\propto t \left( \mu_{\rm A} - \mu_{\rm I} \right)$] in between the two peaks which is characteristic for a Schr\"odinger cat state.}
\label{fig:system}
\end{center}
\end{figure*}

The key property that makes correlated states such as $|{\rm GHZ}\rangle$ useful for enhanced metrology is that they can be thought of as ``bimodal'', in the sense that the probability of an appropriate observable is peaked in two (or more) ``phases'' (the states $|0\rangle^{\otimes N}$ and $|1\rangle^{\otimes N}$ in the case of $|{\rm GHZ}\rangle$).  This bimodality is reminiscent of what occurs near a (first-order) phase transition.  In fact, enhanced parameter estimation can be achieved with pure states at quantum phase transitions \cite{Zanardi}.  For large $N$, highly correlated {\em pure} states are challenging to prepare in practice \cite{preparation}, either as the ground state of a closed many-body system, or as the stationary state of some dissipative dynamics \cite{dissipative}.  Typically, the latter requires careful system engineering, since generic open quantum
systems have {\em mixed} rather than pure stationary states.
In general one therefore has to deal with mixed states.  These however have an additional complication since the best possible measurement is difficult to formulate in general, except for particular cases such as thermal states \cite{thermal}.  This means that with mixed states it is often difficult to compute the best possible precision of parameter estimation.

In this paper we show theoretically how to exploit the dynamics of open quantum systems (for example, driven atomic or molecular ensembles emitting photons \cite{sources}, or quantum dots \cite{qdots}) to generate states for quantum enhanced metrology.  Our approach connects to recent work on parameter estimation with single stationary states of open quantum systems \cite{MadalinLANdiscrete,Molmer,MadalinLANcontinuous}.  
We overcome the problem of mixed states by considering the combined state of the system and output. This is a pure quantum state---actually a matrix product state (MPS) \cite{mps,Lesanovsky2013}---which encodes the state of the system as well as the record of emissions for the whole observation time.  This allows us to find the best estimation precision using the system-output as a resource.

This approach has several advantages. First, it provides improved precision due to the fact that the effective ``size'' of the system and output is now $N t$ where $t$ is the observation time and $N$ is the system size. The second advantage arises from the fact that open systems can feature {\em dynamical} phase transitions (DPTs) \cite{DPT,Ates2012,Lesanovsky2013} which, in contrast to static transitions, are
characterised by singular changes in observables on the whole dynamical evolution and not just on the state of the system.  We show that at a first-order DPT \cite{DPT,Ates2012} the quantum Fisher information (QFI) of the system-and-output may become quadratic in $t$ giving rise to Heisenberg scaling.
We also clarify the behaviour away from the transition point.  Here Heisenberg scaling of the QFI is present 
for times shorter than the correlation time of the dynamics, while asymptotically linear scaling is recovered.  Moreover, due to the pure form of the system-output state we can always (formally) construct the optimal measurement.  We discuss our ideas in a specific setting for quantum enhancement in phase estimation using an intermittent system near a dynamical first-order transition as shown in Fig. \ref{fig:system}(a).

\noindent
{\em Elements of quantum metrology.} We first review some essential aspects of quantum parameter estimation.  Suppose that we wish to estimate a parameter $g$ encoded in a quantum state $\rho_g$, by measuring an observable $M$.  The estimation precision is given by the signal to noise ratio \cite{SNR}, ${\rm SNR}_g(M) =
(d \langle M\rangle_g / dg)^2 / \Delta_g^2 M$, where $\langle M \rangle_g = \mathrm{Tr} (\rho_g M)$ and $\Delta^2_g M = \mathrm{Tr} (\rho_g M^2)-\langle M \rangle_g^2$ are the mean and variance of measuring $M$ on $\rho_g$, respectively.  The observable with the optimal SNR is (up to linear transformations) given by the so-called symmetric logarithmic derivative, ${\cal D}_g$, defined by the relation \cite{QFI}
\begin{equation}
\frac{d\rho_g}{d g} = \frac{1}{2} ({\cal D}_g  \rho_{g} + \rho_{g} {\cal D}_g)  .
\label{eq:log}
\end{equation}
Except for very particular forms of $\rho_{g}$, the optimal measurement ${\cal D}_g$ is difficult to engineer.  Nevertheless, the SNR for this observable is given by the quantum Fisher information~\cite{QFI}, $F(\rho_{g})$, which bounds the precision of any measurement that can be performed in practice.  This bound is in fact given by the variance of ${\cal D}_g$, i.e. $F(\rho_{g})=\Delta^2_g {\cal D}_g$.

In general, the QFI is hard to compute, but for a pure state, $|\psi_g\rangle$, it can be obtained from the fidelity $\langle\psi_{g_{1}}|\psi_{g_{2}}\rangle$ \cite{Zanardi,MadalinLANdiscrete} according to,
\begin{equation}
F(|\psi_{g}\rangle)= \left. 4 \partial^2_{g_{1} g_{2}} \log \langle\psi_{g_{1}}|\psi_{g_{2}}\rangle \right|_{g_{1}=g_{2}=g} .
\label{eq:fidelity1}
\end{equation}
A situation which is relevant for what follows is
when the parameter is encoded as a phase in a unitary transformation on a pure state, $|\psi_g \rangle = e^{-ig\, G} |\psi\rangle$.  Here the fidelity $\langle\psi_{g_{1}}|\psi_{g_{2}}\rangle$ is
the characteristic function of $G$ at $g_{1}-g_{2}$, and the QFI is given by its variance, $F( |\psi_g\rangle) = 4  \Delta^2_{g} G$.  Note that while the QFI is given by the variance of both ${\cal D}_g$ and $G$, these two operators play very different roles.  The optimal measurement to recover the parameter $g$ is ${\cal D}_g$, and its SNR is maximal, ${\rm SNR}_g({\cal D}_g) = F(|\psi_g\rangle)$.  In contrast, $G$ encodes $g$ in the quantum state, but measuring it provides no information about $g$ since ${\rm SNR}_g(G)=0$.  

For example, for the state $|{\rm GHZ}_{g}\rangle$ the generator is $G = \sum_j (1+\sigma_{z}^{(j)})/2$ and the optimal measurement ${\cal D}_g = \sum_j e^{-i g G} \sigma_{y}^{(j)} e^{i g G}$, where $\sigma_{a}^{(j)}$ are Pauli operators acting on qubit $j$. The QFI for the GHZ state then obeys Heisenberg scaling, $F( |{\rm GHZ}_{g}\rangle) = N^2$. This is related to the fact that the distributions of both $G$ and ${\cal D}_g$ are bimodal.  In contrast, the QFI of the uncorrelated state is standard, $F((|0\rangle + e^{-i g}|1\rangle )^{\otimes N})=N$, given by the fact that the corresponding distributions are unimodal.  Below we show that an analogous change from bimodal to unimodal also accompanies a change in the scaling with time of the QFI when approaching a first-order DPT.

\noindent
{\em Open dynamics, MPS and DPTs.}
Our goal is to explore open quantum systems as resources for parameter estimation.
We consider systems whose reduced dynamics, after tracing out the environment, is given by a Markovian master equation \cite{master}
\begin{eqnarray}
\frac{d\rho}{dt}=\mathcal{L} \rho &=& -i[H,\rho] + \sum_{j=1}^k
\left(
L_j \rho L_j^\dagger - \frac{1}{2}\{L_j^\dagger L_j , \rho\}
\right)
\label{eq:master}
\end{eqnarray}
where $H$ is the system's Hamiltonian, and $L_j$ are the jump operators ($j=1,\ldots,k$). In the input-output formalism \cite{GardinerZoller}, the \emph{joint system and output state} is given by a continuous MPS (CMPS) \cite{mps,Lesanovsky2013}. For clarity, we discretise time by $\delta t$, and the CMPS is approximated by a regular MPS \cite{mps,Lesanovsky2013,SM},
$|\Psi(t)\rangle = \sum_{j_n,...,j_1=0}^k   K_{j_n}\cdots K_{j_1}\left|\chi \right> \otimes\left|j_1,..., j_n\right>$,
where $n = t/\delta t$, $K_0=e^{-i \delta t H}\, \sqrt{1-\delta t\sum_j L^\dagger_j L_j}$, $K_{j>0}=e^{-i \delta t H}\,\sqrt{\delta t }L_j$, and $|\chi \rangle$ is the initial state of the system. The output state $|j_1,..., j_n\rangle$ describes the time record of emissions into the environment, as sketched in Fig.\ 1(a).

The state $|\Psi(t)\rangle$ can have a singular change when varying a parameter in \eqref{eq:master}.  This could either correspond to a static phase transition in the stationary state of the system, or to a dynamical phase transition in the system and output.  Both kinds of transitions are captured by discontinuities in the average, or a higher cumulant, of an observable that acts on the whole of $|\Psi(t)\rangle$.

\noindent
{\em Relation between DPTs and QFI.} We now assume that the dynamics depends on the parameter $g$ to be estimated, see Fig.\ 1(a).  This means that the Hamiltonian, $H_{g}$ and jump operators, $L_{j,g}$, and consequently the master operator, $\mathcal{L}_{g}$, Eq.\ (\ref{eq:master}), may depend on $g$.
It follows then that the MPS, $|\Psi_{g}(t)\rangle$, also depends on $g$, and so does the fidelity \cite{SM},
\begin{equation}
\label{eq:fidelity}
\langle \Psi_{g_{1}}(t)|\Psi_{g_{2}}(t)\rangle=\mathrm{Tr} \{ e^{t\mathcal{L}_{g_{1},g_{2}}}|\chi\rangle\langle\chi| \} ,
\end{equation}
where $\mathcal{L}_{g_{1},g_{2}}$ is a deformation of the Master operator \cite{SM},
\begin{eqnarray}
\mathcal{L}_{g_{1},g_{2}} && \rho =
-\,i H_{g_{1}} \rho + i\rho H_{g_{2}} \label{Lgg}\\
&&+\sum_{j=1}^k\left[ L_{j,g_{1}}\rho L_{j,g_{2}}^\dagger-\, \frac{1}{2}( L_{j,g_{1}}^\dagger L_{j,g_{1}} \rho+ \rho L_{j,g_{2}}^\dagger L_{j,g_{2}})\right] \nonumber.
\end{eqnarray}

Thus, in the long time limit the QFI of $|\Psi_{g}(t)\rangle$ is related to the largest eigenvalue $\lambda_{1}(g_{1},g_{2})$ of $\mathcal{L}_{g_{1},g_{2}}$,
\begin{equation}
\label{eq:QFIlongt}
\lim_{t \to \infty} t^{-1} F( |\Psi_{g}(t)\rangle ) = 4 \left.
\partial^2_{g_{1} g_{2}} \lambda_{1}(g_{1},g_{2})
\right|_{g_{1}=g_{2}=g} .
\end{equation}
One can already see that something interesting will occur as the system approaches a DPT, so that the gap between the two leading eigenvalues of $\mathcal{L}_{g}$ closes at some $g$.  

When the gap is small, for example close to a DPT, there is a time regime where the QFI is quadratic in time,
\begin{eqnarray}
&& F( |\Psi_{g}(t)\rangle ) = 4 t^{2}
\partial^2_{g_{1}g_{2}}
\mathrm{Re}\, \mathrm{Tr}
\left\{
\mathcal{L}_{g_{1},g} {\mathcal P}   \mathcal{L}_{g,g_{2}} {\mathcal P} |\chi\rangle\langle\chi|
\right\}_{g_{1}=g_{2}=g}
\nonumber \\
&& \label{Ft2}
\;\;
- \left|
2 t
\partial_{g_{1}}
\mathrm{Tr} \left\{\mathcal{L}_{g_{1},g} {\mathcal P} |\chi\rangle
\langle\chi|
\right\}_{g_{1}=g}
\right|^{2} +
t^{2} {\mathcal O}(t \lambda_{2}) + {\mathcal O}(t) ,
\end{eqnarray}
where ${\mathcal P}$ is a projection onto the first two eigenvectors of $\mathcal{L}_{g}$ corresponding to the two eigenvalues with the largest real part, $(\lambda_{1}=0, \lambda_{2})$. The gap is given by $- \mathrm{Re} \; \lambda_{2}$.  This approximation of Eq.\ \eqref{Ft2} is valid for $\tau' \ll t \ll \tau$, where $\tau$ is the correlation time given by the gap, $\tau \equiv (- \mathrm{Re} \; \lambda_{2})^{-1}$, while $\tau'$ is the longest timescale associated with the rest of the spectrum, $\tau' \equiv (- \mathrm{Re} \; \lambda_{3})^{-1}$.  The quadratic time dependence of the QFI \eqref{Ft2} is a consequence of time-correlations in the system-output MPS.  Furthermore, at a DTP $\lambda_{2} \to 0$ and the asymptotic scaling of Eq.\ \eqref{eq:QFIlongt} is no longer valid.  Instead the QFI is quadratic in time and this {\em Heisenberg scaling} is given exactly by the Eq.\ \eqref{Ft2} for all $t \gg \tau'$.

\noindent
{\em Enhanced phase estimation and intermittency.} We now use the ideas above
for the case of a system with intermittent dynamics
\cite{qdots,Plenio1998,Ates2012}
used as a resource for parameter estimation, see Fig.\ 1(a).  The parameter here is a phase $g=\phi$ encoded in the jump operator $L_{1}$, by defining $L_{1,\phi} = e^{-i \phi} L_{1}$.  For concreteness, note that the quantum jump associated with $L_{1}$ is the emission of a photon.  This means that a phase $\phi$ is imprinted on each outgoing photon.
As we now show, if the system displays intermittent photon emission associated to a (first-order) DPT in counting statistics \cite{DPT,Ates2012,Lesanovsky2013}, then it will be an efficient resource for quantum metrology.
With the above choice, the Master operator is independent of $\phi$, $\mathcal{L}_{\phi}=\mathcal{L}$.  In turn, the deformed generator $\mathcal{L}_{\phi,\phi'}$, Eq.\ \eqref{Lgg}, from which the QFI is obtained, reads ($\Delta \phi=\phi-\phi'$)
\begin{equation}
\mathcal{L}_{\phi,\phi'} \rho
= \mathcal{L} \rho +
\left(e^{-i\Delta \phi} - 1 \right) L_{1} \rho L_{1}^\dagger .
\label{Lphi}
\end{equation}

With these definitions there is a direct connection to a photon counting problem \cite{DPT,GardinerZoller}.  The phase $\phi$ is encoded in a unitary transformation of the MPS with generator $G=\Lambda(t)$, where $\Lambda(t)$ is the operator that counts the number of photons emitted up to time $t$, so that $|\Psi_\phi(t)\rangle = e^{-i \phi \Lambda(t)} |\Psi(t)\rangle$.  The fidelity $\langle \Psi_\phi(t)|\Psi_{\phi'}(t)\rangle$ is the characteristic function of $\Lambda(t)$, the logarithm of which encodes all its cumulants.  The cumulants are also encoded in the cumulant generating function (CGF), $\Theta_{t}(s) = \log \sum_{\Lambda} e^{-s \Lambda} P(\Lambda,t)$, where
$P(\Lambda,t)$ is the probability of observing $\Lambda$ photons in time $t$. The CGF can be related to a deformation of the Master operator, $\Theta_{t}(s) = \mathrm{Tr} \{ e^{t \mathcal{L}(s)} |\chi\rangle \langle\chi| \}$, where $\mathcal{L}(s)$ is the same as \eqref{Lphi} with $\Delta \phi = - i s$.  The long time limit of the CGF, $\theta(s) = \lim_{t \to \infty} t^{-1} \Theta(s,t)$, plays the role of a dynamical free-energy for the ensemble of trajectories of photon emissions \cite{DPT}.  A singularity of $\theta(s)$ at some $s_{c}$ is an indication of a phase transition in the ensemble of quantum jump trajectories, and when $s_{c}=0$ we have what we term a DPT, i.e., a singular change in the actual dynamics of the open system associated with a vanishing of the spectral gap $\lambda_{2}$ \cite{DPT,Lesanovsky2013}.

The asymptotic QFI \eqref{eq:QFIlongt} becomes,
\begin{equation}
\label{eq:QFIlongtcount}
\lim_{t \to \infty} t^{-1} F( |\Psi_{g}(t)\rangle ) = 4 \left.
\partial_s^2\theta(s)
\right|_{s=0} .
\end{equation}
When the function $\theta(s)$ has a first-order singularity at some $|s_{c}| \gtrsim 0$, i.e.\ we are near a DPT, Eq.\  \eqref{eq:QFIlongtcount} will be large at $s=0$.  In this a case the system will display an intermittent dynamics that switches between long periods with very distinct emission characteristics.  Such a situation can be understood in terms of the coexistence of dynamical phases with significantly different photon count rates \cite{DPT}, see Fig.\ 1(a).  The QFI of $|\Psi_\phi(t)\rangle$ is proportional to the variance of the photon counting generator $G=\Lambda(t)$.
For times shorter than the correlation time $\tau$ the system is mostly in one of the two phases, the distribution of the photon count is approximately bimodal, and the dynamics displays large fluctuations in the total photon emission, see Fig.\ 1(b).
This implies a quadratic increase of the QFI with time, with Eq.\ \eqref{Ft2} reducing to,
\begin{equation}
F( |\Psi_{g}(t)\rangle ) \approx 4 t^{2} p_{\rm A} p_{\rm I} \left( \mu_{\rm A} - \mu_{\rm I} \right)^{2} + \mathcal{O}(t).
\label{Ft2count}
\end{equation}
Here $\mu_{\rm A}$ and $\mu_{\rm I}$ are the average counting rates, $\langle \Lambda(t) \rangle / t$, in the two phases (which we term ``active'' and ``inactive'' as we assume $\mu_{\rm A} > \mu_{\rm I}$), while $p_{\rm A}$ and $p_{\rm I}$ are the probability of the initial state $|\chi\rangle$ being in either phase.  The above approximation holds for $t < \tau$, and becomes valid for all times at a DPT.
For times longer than $\tau$, dynamics switches between the two phases, giving rise to intermittent behaviour, and eventual normal (unimodal) distribution of the photon count around the overall average \eqref{eq:QFIlongtcount}; see Fig.\ 1(b) and derivations in \cite{SM}.

The above shows that an intermittent system near a DPT can be used as a photon source for quantum enhanced phase estimation.    The situation is then similar to that of GHZ states: the total photon count distribution is bimodal for times up to the correlation time $\tau$ and imprints an effective macroscopic phase difference of $t \left( \mu_{\rm A} - \mu_{\rm I} \right) \phi$ between the active and inactive dynamical phases; see discussion after Eq.\ \eqref{eq:coh}.

\noindent
{\em Enhanced metrology and DPT in general.}
We now extend the above discussion to the case where the dynamics has an arbitrary dependence on the parameter $g$ to be estimated.  In this case, $g$ is encoded in the action of a ``generator" $G_g(t)$,
\begin{equation}
\label{generatorGEN}
G_g(t)\,|\Psi_g (t) \rangle =
-i \partial_g |\Psi_g(t) \rangle,
\end{equation}
where $G_g(t)$ is the time-integral of a local-in-time observable, just like $\Lambda(t)$ in the photon counting case.  In terms of $G_g(t)$ the fidelity reads, $\langle \Psi_{g_{1}}(t)| \Psi_{g_{2}}(t)\rangle
=
\langle \Psi_{g}(t)|  \mathcal{T} e^{ - i\int_{g_{1}}^{g_{2}} dg'\, G_{g'}(t)}\,| \Psi_{g}(t)\rangle$,
where $\mathcal{T}$ is the $g$-ordering (cf.\ time-ordering) operator, see also~\cite{AnnabellaUnitaryDisturbance,MadalinJukka}.  The QFI is then the variance of $G_g(t)$.
It follows that if we have a system which displays a first-order DPT where the dynamical phases are characterised by $G_g(t)$,
then, in the $\tau' \ll t \ll \tau$ time regime, the QFI follows Eq.\ \eqref{Ft2count},
where $\mu_{\rm A,I}$ are the averages of $G_g(t)$ per unit time in the two coexisting dynamical phases \cite{NextPaper}.  Again this emphasises the connection between dynamical bimodality and enhanced quantum sensitivity.

The $t^{2}$ behaviour of the QFI is an intrinsically quantum feature.
This behaviour cannot occur in systems for which the associated MPS is real and therefore cannot accumulate any quantum phase.  Note that this includes all classical systems.
In such a case the average of $G_{g} (t)$ is zero, cf.\ Eq.\ \eqref{generatorGEN}, and only terms linear in $t$ will survive in Eq.\ \eqref{Ft2count}.

\noindent
{\em Measurement schemes.} We have shown that near a DPT the system-output state can have a large QFI.  But to exploit this, and achieve quantum enhanced sensitivity, it is necessary to measure an appropriately chosen observable.  The optimal observable is known to be the symmetric logarithmic derivative ${\cal D}_g$ defined by (\ref{eq:log}), which for pure states can be written explicitly as ${\cal D}_g=2 \partial_g |\psi_g\rangle \langle\psi_g|$. However, the measurement of ${\cal D}_g$ will be  difficult to engineer in most practical situations.  One needs therefore to find an alternative which is both practical and whose SNR is as close as possible to the QFI.

Despite the fact that the intricacy of the optimal measurement makes it impractical, we can still formulate general characteristics for a measurement that achieves enhanced precision.  The first consideration is whether the measurement should be on the system or output, or both.  In fact, in the regime of enhanced scaling the optimal measurement whose precision is given by the QFI involves measuring both system and output. The reason is that the precision achievable by measuring only the output is bounded by $p_{\rm A} F(|\Psi_{\rm A} (t)\rangle)+ p_{\rm I} F(|\psi_{\rm I} (t)\rangle)$,
which scales linearly in time. Here $|\Psi_{\rm A,I} (t)\rangle$ are the MPS states associated to the individual active/inactive stationary states, and $p_{\rm A,I}$ are their probabilities, see Eq.\ \eqref{Ft2count}.  This last result is the precision of an idealised protocol given by a
first measurement of the system to project onto one of the subspaces associated with the competing stationary states, followed by an optimal measurement of the conditioned system-output state $|\Psi_{\rm A,I} (t)\rangle$. The second consideration is what should be the time extension $t$ of a single measurement run.
Here we imagine that the total time available to the experiment is $T$ and one performs $n=T/t$ independent repetitions of an efficient system-output measurement of the state $|\Psi_g(t)\rangle$.  
This corresponds to a measurement of the joint state $|\Psi_g(t)\rangle^{\otimes n}$, and the optimal time $t$ 
is that which maximises the QFI of the joint state, $F(|\Psi_g(t)\rangle^{\otimes n}) = n \, F(|\Psi_g(t)\rangle)$.  Equation \eqref{Ft2} tells us that this optimal time is of the order of the correlation time, $t = {\cal O}(\tau)$.

For the case of phase estimation at a DPT, the bimodality of the system-output state means that it is essentially of the form of a
``Schr\"{o}dinger cat'' state.  Assuming for simplicity that the competing stationary states are pure, it reads,
\begin{equation}
|\Psi_\phi(t)\rangle =
\sqrt{p_{\rm I}} \,
| {\rm I} \rangle \otimes |\alpha_{\rm I}(\phi)\rangle +
\sqrt{p_{\rm A}} \,
| {\rm A} \rangle \otimes |\alpha_{\rm A}(\phi)\rangle
\label{eq:coh}
\end{equation}
where $|\alpha_{\rm A}(\phi)\rangle$ are coherent states with amplitudes $\alpha_{\rm I,A}(\phi) = e^{i \phi} \sqrt{t \, \mu_{\rm I,A}}$, where $\mu_{\rm I,A}$ are the photon emission rates of the phases,
see Eq.\ \eqref{Ft2count} and Fig.\ 1(c).
In fact, as shown in Ref.\ \cite{Ralph}, the state
\eqref{eq:coh} is approximately a GHZ state with relative phase $t (\mu_{\rm A}-\mu_{\rm I}) \phi$.
Note that for \eqref{eq:coh} neither counting nor homodyne measurements achieve Heisenberg scaling, which highlights the general challenge of identifying optimal measurements. However, one might think of instead employing interferometric protocols, related to the ones put forward in Refs.\ \cite{Ralph,JooMunroSpiller,Gerry} for superpositions of coherent states, in order to exploit the enhanced precision scaling.

\noindent
{\em Conclusions.} We have shown that, close to a dynamical phase transition,
the output of an open quantum system can be seen as a resource for quantum metrology applications.
For times of the order of the correlation time, the system-output QFI scales quadratically with time, while in the long time limit the QFI scales linearly in time with rate which diverges when the spectral gap closes, as in a DTP.
It remains an open issue how to exploit
in a general and systematic way
the large QFI of the system-output close to a DPT.

\acknowledgments
This work was supported by The Leverhulme Trust (Grant No.\ F/00114/BG), EPSRC (Grant No.\ EP/J009776/1) and the European Research Council under the European Union's Seventh Framework Programme (FP/2007-2013) through ERC Grant Agreement No.\ 335266 (ESCQUMA) and the EU-FET Grant No.\ 512862 (HAIRS). I.L. acknowledges discussions with K. M{\o}lmer and E.M. Kessler.

\widetext
\newpage

\section{SUPPLEMENTARY MATERIAL}
\section{I. \,Fidelity and QFI of MPS states}

In this section we prove Eqs. (2) and (4) from the paper.

We have:
\begin{equation}
\partial^2_{g_1g_2}\log\langle\psi_{g_1}|\psi_{g_2}\rangle|_{g_1=g_2=g} \,=\,  \frac{\langle \psi_{g}'|\psi_{g}'\rangle}{\langle \psi_{g}|\psi_{g}\rangle} -  \frac{\langle \psi_{g}'  |\psi_{g}\rangle \langle \psi_{g}  |\psi_{g}'\rangle }{\langle \psi_{g}|\psi_{g}\rangle^2}= \langle \psi_{g}'|\psi_{g}'\rangle -  \left|\langle \psi_{g}  |\psi_{g}'\rangle \right|^2
\end{equation}
where $|\psi_{g}'\rangle=\frac{\partial}{\partial g_1}|_{g_1=g}|\psi_{g_1}\rangle$ and we use the normalisation of the state $\langle\psi_{g}|\psi_{g}\rangle=1$.

On the other hand, for a family of pure states $\rho_g=|\psi_g\rangle\langle\psi_g|$ the symmetric logarithmic derivative is $\mathcal{D}_g=2(|\psi_g\rangle\langle\psi_g'| +|\psi_g'\rangle\langle\psi_g|)$.
Therefore
\begin{eqnarray}
F(|\psi_g\rangle)=\mathrm{Tr}(\rho_g \mathcal{D}_g^2)&=& 4 ( \langle \psi_{g}'|\psi_{g}'\rangle + \langle \psi_{g}'  |\psi_{g}\rangle \langle \psi_{g}  |\psi_{g}'\rangle+ \langle \psi_{g}'  |\psi_{g}\rangle^2+\langle \psi_{g}  |\psi_{g}'\rangle^2) \nonumber\\
&=& 4 ( \langle \psi_{g}'|\psi_{g}'\rangle - \langle \psi_{g}'  |\psi_{g}\rangle\langle \psi_{g}  |\psi_{g}'\rangle + \left(\langle \psi_{g}'  |\psi_{g}\rangle+ \langle \psi_{g}  |\psi_{g}'\rangle\right)^2)   \nonumber\\
&=& 4 ( \langle \psi_{g}'|\psi_{g}'\rangle - |\langle \psi_{g}'  |\psi_{g}\rangle|^2)  = 4\,\partial^2_{g_1g_2}\log|\langle\psi_{g_1}|\psi_{g_2}\rangle|_{g_1=g_2=g} .
\end{eqnarray}
In order to prove Eq. (4), let us consider the discretisation of the master dynamics described below Eq. (3) of the paper. We have:
\begin{equation}
\langle\Psi_{g_1}(t)|\Psi_{g_2}(t)\rangle=\mathrm{Tr}\left\{|\Psi_{g_2}(t)\rangle \langle\Psi_{g_1}(t)|\right\}=\mathrm{Tr}\left\{\sum_{j_n,...,j_1=0}^k  K_{j_n,g_2}\cdots K_{j_1,g_2}\left|\chi \left\rangle\right\langle\chi\right| K^\dagger_{j_n,g_1}\cdots K^\dagger_{j_1,g_1}\right\},
\end{equation}
where $n = t/\delta t$, $K_{0,g}=e^{-i \delta t H_g}\, \sqrt{1-\delta t\sum_{j=1}^k L^\dagger_{j,g} L_{j,g}}$, $K_{j>0,g}=e^{-i \delta t H_g}\,\sqrt{\delta t }L_{j,g}$, and $|\chi \rangle$ is the initial state of the system. In the limit $\delta t \to 0$, analogously as tracing out the output in $|\Psi_g(t)\rangle $ gives the state of the system $\rho_g(t)$: 
\begin{equation*}
\rho_g(n)=\sum_{j_n,...,j_1=0}^k  K_{j_n,g}\cdots K_{j_1,g}\left|\chi \left\rangle\right\langle\chi\right| K^\dagger_{j_n,g}\cdots K^\dagger_{j_1,g}\quad\underset{\delta t\rightarrow 0}{\longrightarrow} \quad \rho_g(t)=e^{t\mathcal{L}_g}\left|\chi \rangle\langle\chi\right|,
\end{equation*}
 the fidelity becomes $\langle\Psi_{g_1}(t)|\Psi_{g_2}(t)\rangle =\mathrm{Tr}\{e^{t\mathcal{L}_{g_1,g_2}}\left|\chi \rangle\langle\chi\right|\}$, where $\mathcal{L}_{g_1,g_2}$ is a modified Master operator defined in Eq. (5) in the paper. The same result can also be derived by using the continuous MPS state which describes the state of the system and the output in continuous time:
\begin{eqnarray}
|\Psi(t)\rangle\,&&=\sum_{m=0}^\infty\sum_{j_1,...,j_m=1}^k \int_{0}^t \mathrm{d}t_1 \int_{t_1}^t\mathrm{d}t_2\cdots\, \int_{t_{m-1}}^t\mathrm{d}t_m \nonumber\\
&&\,\times\,\left(e^{-i (t-t_{m})H^{\rm eff}}  L_{j_m} e^{-i (t_m-t_{m-1})H^{\rm eff}} \cdots\,L_{j_2} e^{-i (t_2-t_1)H^{\rm eff}} L_{j_1} e^{-i t_1 H^{\rm eff}}\, |\chi\rangle\right) \otimes|(j_1,t_1),(j_1,t_2),...,(j_m,t_m)\rangle
\end{eqnarray}
where $H^{\rm eff}=H-i\sum_{j=1}^k L_j^\dagger L_j$ is the effective Hamiltonian.

\section{II.\, Time dependence of QFI}

In this section we first discuss the general dependence of the QFI of the MPS state $|\Psi(t)\rangle$ on time $t$. This will enable us to prove the asymptotic linear behaviour of the QFI  in the case of dynamics with a unique stationary state, see Eq. (6) in the paper. Using the general time dependence of the QFI, we then prove the existence of a quadratic scaling regime of the QFI (cf. Eq. (7) in the paper) for dynamics near a DPT. Finally, for a system displaying a first-order DPT in photon emissions, we argue how the quadratic scaling of the QFI for phase estimation with emitted photons, can be related to difference in photon emission rates between dynamical phases, cf. Eq. (10) in the paper.

\subsection{A. \,General time dependence of the QFI} In order to express the QFI of the MPS state $|\Psi(t)\rangle$, we use Eqs. (2) and (4) in the paper, and  obtain:
\begin{eqnarray}
F(|\Psi_g(t)\rangle )&=& 4 \, \partial^2_{g_1 g_2} \log \mathrm{Tr}\, \{e^{t\mathcal{L} _{g_1,g_2}}|\chi\rangle\langle\chi|\}_{g_1=g_2=g} = - 4 \left|\mathrm{Tr}\, \left\{\int_{0}^t \mathrm{d}t'\, \partial_{g_1} \mathcal{L} _{g_1,g} \,\rho_g(t')\right\}_{g_1=g}\right|^2 \nonumber\\
&&  + 4\, \mathrm{Tr}\, \left\{\int_{0}^t \mathrm{d}t'\, \partial^2_{g_1 g_2} \mathcal{L} _{g_1,g_2} \,\rho_g(t')\right\}_{g_1=g_2=g}\nonumber\\
&&  + 8\, \mathrm{Re}\, \mathrm{Tr}\, \left\{\int_{0}^t \mathrm{d}t'\, \int_{0}^{t-t'} \mathrm{d}t''\, \partial_{g_1} \mathcal{L} _{g_1,g} e^{t'' \mathcal{L} _{g}}\, \partial_{g_2} \mathcal{L} _{g,g_2} \,\rho_g(t')\right\}_{g_1=g_2=g}\label{sm:QFItime}
\end{eqnarray}
where $\rho_g(t):=e^{t\mathcal{L}_g}|\chi\rangle\langle\chi| $, $\mathcal{L}_g$ is the Master operator, see Eq. (3) in the paper, and $\mathcal{L}_{g_1,g_2}$ is the modified Master operator, see Eq. (5) in the paper. 


For clarity of further presentation we assume that $\mathcal{L} _{g}$ can be diagonalised and has one stationary state $\rho_{ss}$, i.e $\mathcal{L}_g=0|\rho_{ss}\rangle\langle1|+\sum_{k=2}^{d^2}\lambda_k |\rho_{k}\rangle\langle l_k|$, where $d$ is the dimension of the system Hilbert space $\mathcal{H}$ and $|\rho_k\rangle$, $\langle l_k|$ stand for $k$-th right and left eigenvectors of $\mathcal{L}_g$, ordered so that $0>\mathrm{Re}\,\lambda_2\geq\mathrm{Re}\,\lambda_3\geq...\geq\mathrm{Re}\,\lambda_{d^2}$ and normalised as $\langle l_j|\rho_k\rangle=\delta_{jk}$, $j,k=1,...,d^2$. The stationary state $\rho_{ss}$ and other eigenvectors $\{|\rho_k\rangle,\langle{l_k}|\}_{k\geq2}^{d^2}$ with corresponding eigenvalues $\{\lambda_k\}_{k\geq2}^{d^2}$ depend on $g$. In general one should consider a Jordan decomposition of $\mathcal{L}_{g}$, but the following discussion would be similar for that case.

Due to the fact that Eq.\ \eqref{sm:QFItime} involves integrals of $e^{t\mathcal{L}_g}$, we need to consider the $0$-eigenspace of $\mathcal L_g$, i.e. the stationary state $\rho_{ss}$, separately from all the rest of eigenvectors, whose eigenvalues are different from 0. We introduce the projection on the complement of the stationary state $\mathcal{P}_1:=\sum_{k=2}^{d^2}|\rho_k\rangle\langle l_k|$ and denote the restriction of an operator $\mathcal{X}$ to the complement of $\rho_{ss}$ by $[\mathcal{X}]_{\mathcal{P}_1}:=\mathcal{P}_1 \mathcal{X} \mathcal{P}_1$. 
 
We now express the finite time behaviour of QFI using derivatives of the modified Master operator 
and the diagonal decomposition of the original Master operator $\mathcal{L}_{g}$. From Eq.\ \eqref{sm:QFItime} it follows:
\begin{eqnarray}
F(|\Psi_g(t)\rangle )&=& -4\left|\,t\, \mathrm{Tr}\,\left\{\partial_{g_1} \mathcal{L} _{g_1,g} \,\rho_{ss}\right\}+\mathrm{Tr}\,\left\{\partial_{g_1} \mathcal{L} _{g_1,g}\left[\frac{e^{t\mathcal{L}_g}-\mathcal{I}}{\mathcal{L}_g} \right]_{\mathcal{P}_1}|\chi\rangle\langle\chi|\right\}\right|_{g_1=g}^2 \nonumber\\
&&  + 4\,  \left(\,t\,\mathrm{Tr}\,\left\{\partial^2_{g_1g_2} \mathcal{L} _{g_1,g_2} \,\rho_{ss}\right\}+\mathrm{Tr}\,\left\{\partial^2_{g_1g_2} \mathcal{L} _{g_1,g_2}\left[\frac{e^{t\mathcal{L}_g}-\mathcal{I}}{\mathcal{L}_g} \right]_{\mathcal{P}_1}|\chi\rangle\langle\chi|\right\}\right)_{g_1=g_2=g}\nonumber\\
&&  + 4\, t^2 \,\left|\mathrm{Tr}\,\left\{\partial_{g_1} \mathcal{L} _{g_1,g} \,\rho_{ss}\right\}\right|^2 +8 \,\mathrm{Re}\, \mathrm{Tr}\,\left\{\partial_{g_1} \mathcal{L} _{g_1,g} \,\rho_{ss}\right\}\mathrm{Tr}\,\left\{\partial_{g_2} \mathcal{L} _{g,g_2}\left[\frac{e^{t\mathcal{L}_g}-\mathcal{I}-t\mathcal{L}_g}{\mathcal{L}_g^2}  \right]_{\mathcal{P}_1}|\chi\rangle\langle\chi|\right\}_{g_1=g_2=g} \nonumber\\
&&  + 8 \,\mathrm{Re} \,\mathrm{Tr}\,\left\{\partial_{g_1} \mathcal{L} _{g_1,g}\left[  \frac{e^{t\mathcal{L}_g}-\mathcal{I}-t\mathcal{L}_g}{\mathcal{L}_g^2} \, \right]_{\mathcal{P}_1}\partial_{g_2} \mathcal{L} _{g,g_2}\, \rho_{ss} \right\}_{g_1=g_2=g}
\nonumber\\  
&& -8 \,\mathrm{Re} \,\mathrm{Tr}\,\left\{
\partial_{g_1} \mathcal{L} _{g_1,g} \left[\mathcal{L}_g^{-1} \right]_{\mathcal{P}_1}\partial_{g_2} \mathcal{L} _{g,g_2}\left[ \frac{e^{t\mathcal{L}_g}-\mathcal{I}}{\mathcal{L}_g} \right]_{\mathcal{P}_1}|\chi\rangle\langle\chi|\right\}_{g_1=g_2=g}\nonumber\\
&& +  8 \,\mathrm{Re} \,\mathrm{Tr}\,\left\{\partial_{g_1} \mathcal{L} _{g_1,g}\left[  \frac{e^{t\mathcal{L}_g}}{\mathcal{L}_g} \, \left(\int_{0}^{t} \mathrm{d}t'\, e^{-t' \mathcal{L}_{g}}\, \partial_{g_2} \mathcal{L}_{g,g_2} \, e^{t' \mathcal{L} _{g}}\right) \right]_{\mathcal{P}_1}|\chi\rangle\langle\chi|\right\}_{g_1=g_2=g}, \label{sm:QFItime2}
\end{eqnarray}
and one can show that
\begin{eqnarray*}\left[  \frac{e^{t\mathcal{L}_g}}{\mathcal{L}_g} \, \left(\int_{0}^{t} \mathrm{d}t'\, e^{-t' \mathcal{L}_{g}}\, \partial_{g_2} \mathcal{L}_{g,g_2} \, e^{t' \mathcal{L} _{g}}\right) \right]_{\mathcal{P}_1}&=&
\, t\,\sum_{k=2}^{d^2}\frac{e^{t\lambda_k}}{\lambda_k}\langle l_k | \partial_{g_2} \mathcal{L} _{g,g_2}  |\rho_k\rangle \, |\rho_k\rangle \langle l_k |\\
&&+\ \sum_{j\neq k, j,k>1}^{d^2} \frac{e^{t \lambda_j }-e^{t\lambda_k}}{\lambda_k(\lambda_j-\lambda_k) } \langle l_k | \partial_{g_2} \mathcal{L} _{g,g_2}  |\rho_j\rangle \, |\rho_k\rangle \langle l_j |.
\end{eqnarray*} 
The first line and the second line in  Eq.\ \eqref{sm:QFItime2} correspond to the first and the second line in Eq.\ \eqref{sm:QFItime}, respectively. All other terms  in  Eq.\ \eqref{sm:QFItime2} correspond to the third line in Eq.\ \eqref{sm:QFItime}. We see that the quadratic contribution $t^2 \,\left|\mathrm{Tr}\,\left\{\partial_{g_1} \mathcal{L} _{g_1,g} \,\rho_{ss}\right\}\right|^2$ cancels out and for one stationary state there is no explicit quadratic behaviour. 

Eq.\ \eqref{sm:QFItime2} will be used for investigating the asymptotic and the quadratic time regime of QFI.\\

We note that as an alternative route, one can use the eigendecomposition of the modified Master operator $\mathcal{L}_{g_1,g_2}$  defined in Eq. (5) in the paper:
\begin{eqnarray*}
&&e^{t\mathcal{L}_{g_1,g_2}}=\sum_{k=1}^{d^2}e^{t\lambda_k(g_1,g_2)}\,   |\rho_k(g_1,g_2)\rangle\langle l_k(g_1,g_2)|,\quad\mathrm{which\,gives}\\
 &&\mathrm{Tr}\, \{e^{t \mathcal{L} _{g_1,g_2}}|\chi\rangle\langle\chi|\}=\sum_{k=1}^{d^2} p_k (g_1,g_2) e^{t\lambda_k(g_1,g_2)},
\end{eqnarray*}
where $p_k(g_1,g_2)=\left.\langle l_k(g_1,g_2) \right||\chi\rangle\langle\chi|\rangle \times \mathrm{Tr}\{\rho_k(g_1,g_2)\}$ and in the case of a single stationary state we have $p_1(g,g)=1$, $p_k(g,g)=0$, $k=2,...d^2$, which follows from  the normalisation of the eigenbasis of $\mathcal{L}_g$ to $\mathrm{Tr}\{ \rho_1(g,g)\}=\mathrm{Tr} \rho_{ss}=1$, i.e. 
$\mathrm{Tr} \{\rho_k(g,g)\}=0$ for $k=2,...,d^2$. 

From Eqs. (2) and (4) in the paper, for a single stationary state, we obtain:
\begin{eqnarray}
&&F(|\Psi_g(t)\rangle )=\nonumber \\
&&  -4\left (  t^2 \, |\partial_{g_1} \lambda_1(g_1,g)|^2  + 2 \,t \,\mathrm{Re} \, \partial_{g_1} \lambda_1(g_1,g) \sum_{k=1}^{d^2} e^{t \lambda_k}\partial_{ g_2}\,p_k(g,g_2) +   \sum_{j,k=1}^{d^2} e^{t (\lambda_k+\lambda_j)}\partial_{ g_1}\,p_j(g_1,g)\,\partial_{ g_2}\,p_k(g,g_2)\right)_{g_1=g_2=g_0} \nonumber\\
&&+ 4 \left(t^2 \, |\partial_{g_1} \lambda_1(g_1,g)|^2 + t\, \partial^2_{g_1 g_2} \lambda_1(g_1,g_2)+ \sum_{k=1}^{d^2} e^{t \lambda_k}\partial^2_{ g_1g_2}\,p_k(g_1,g_2) +2 \,t\,\mathrm{Re}  \sum_{k=1}^{d^2}  e^{t \lambda_k} \partial_{ g_1} p_k(g_1,g)\, \partial_{ g_2} \lambda_k(g,g_2)  \right)_{g_1=g_2=g_0}\label{sm:QFItimeZ}
\end{eqnarray}
where $\lambda_k=\lambda_k(g,g)$ and $p_k=p_k(g,g)$, $k=1,...,d^2$. The first line correponds to the first line of Eq.\ \eqref{sm:QFItime2} and the second to the rest of terms in Eq.\ \eqref{sm:QFItime2}. We see again that quadratic terms  $t^2 \, |\partial_{g_1} \lambda_1(g_1,g)|^2$ cancel out and there is no explicit quadratic behaviour.\\

\subsection{B. \, Asymptotic QFI for the case of a unique stationary state}

Here we assume that the dynamics has a unique stationary state, i.e. the second eigenvalue of the Master operator $\mathcal{L}_g$ is different from 0, $\lambda_2\neq0$. In order to find the asymptotic behaviour of the QFI of the state $|\Psi_g(t)\rangle$, we consider the limit $t\rightarrow\infty$ when we have $\lim_{t\rightarrow \infty}\left[e^{t\mathcal{L}_{g}}\right]_{\mathcal{P}_1}=0$ and from Eq.\ \eqref{sm:QFItime2} we obtain:
\begin{equation}
\lim_{t\rightarrow\infty} t^{-1} F(|\Psi_g(t)\rangle )=4\, \mathrm{Tr}\,\left\{\partial^2_{g_1g_2} \mathcal{L} _{g_1,g_2} \,\rho_{ss}\right\} - 8 \,\mathrm{Re} \,\mathrm{Tr}\,\left\{\partial_{g_1} \mathcal{L} _{g_1,g} \left[  \mathcal{L}_g^{-1}\right]_{\mathcal{P}_1}\partial_{g_2} \mathcal{L} _{g,g_2}\, \,\rho_{ss}\right\}_{g_1=g_2=g},
\label{sm:QFItimeAlinear}
\end{equation}
Since the limit is finite, this shows that the QFI has an asymptotic linear behaviour in the case of a single stationary state. This result was also obtained using different methods in~\cite{MadalinLANcontinuous}. We see that Eq.\ \eqref{sm:QFItimeAlinear} can diverge at a first-order DPT when $\lambda_2\rightarrow0$ for $g\rightarrow g_c$, as $\left[\mathcal{L}_{g}^{-1}\right]$ has then a diverging eigenvalue $\lambda_2^{-1}$. 

The asymptotic linear behaviour of the QFI can be also obtained from Eq. \eqref{sm:QFItimeZ}:
\begin{equation}
\lim_{t\rightarrow\infty} t^{-1} F(|\Psi_g(t)\rangle )= \partial^2_{g_1g_2}\lambda_1(g_1,g_2)|_{g_1=g_2=g},\label{sm:QFItimeAlinearZ},
\end{equation}
which corresponds to Eq. (6) in the paper. By comparing it to Eq.\ \eqref{sm:QFItimeAlinear}, we see that closing of the gap $\lambda_2\rightarrow0$ when approching $g\rightarrow g_c$, can cause non-analyticity of the eigenvalues and eigenvectors of $\mathcal{L}_{g_1,g_2}$ at $g_1=g_2=g_c$.

\subsection{C. \,Quadratic time-regime of QFI}

In this subsection we describe the quadratic regime in the QFI scaling with time, which can be present for systems at and near a first-order DPT, and is given by Eq. (7) in the paper.  \\

\emph{Quadratic behaviour near a DPT}. We consider a system near a DPT when the gap is much smaller than the gap associated with the rest of the spectrum, i.e. $(-\mathrm{Re}\lambda_2)\ll (-\mathrm{Re}\lambda_3)$. This allows to introduce a time regime $(-\mathrm{Re}\lambda_3)^{-1}=\tau'\ll t\ll \tau=(-\mathrm{Re}\lambda_2)^{-1}$. For simplicity let us assume $|\lambda_2|\approx 0$, which gives $e^{\lambda_2 t}\approx 1$. We introduce the projection $\mathcal{P}:=|\rho_{ss}\rangle\langle 1|+|\rho_{2}\rangle\langle l_2|$  on the subspace spanned by the stationary state $\rho_{ss}$ and the second eigenvector $\rho_2$. We also introduce the projection on their complement $\mathcal{P}_2:=\mathcal{I-P}=\sum_{k=3}^{d^2}|\rho_{k}\rangle\langle l_k|$ and denote by $\left[X \right]_{\mathcal{P}_2}=(\mathcal{I-P})X (\mathcal{I-P})$  the restriction of an operator $X$ to this complement. 

We would like to simplify Eq.\ \eqref{sm:QFItime2} in the regime $\tau'\ll t\ll \tau$. In this regime we expect the second eigenvector $\rho_2$ of $\mathcal{L}_g$ to be almost stationary and determine, both with the stationary state $\rho_{ss}$, dominant terms in the behaviour of the QFI. We assume that other eigenvectors do not play significant role, by which we understand that $\left\lVert \left[e^{t \mathcal{L}_g}\right]_{\mathcal{P}_2}\right\rVert_1\approx0$, where $\left\lVert \tau\right\rVert_1:=\mathrm{Tr}\{\sqrt{\tau^\dagger\tau}\}$ is the trace-norm of operators on $\mathcal{H}$, which for density matrices is always $1$. 

The general behaviour of the QFI given by Eq.\ \eqref{sm:QFItime2} simplifies to: 
\begin{eqnarray}
F(|\Psi_g(t)\rangle )&=& -4\left|\,t\, \mathrm{Tr}\,\left\{\partial_{g_1} \mathcal{L} _{g_1,g}\,\mathcal{P} \,|\chi\rangle\langle\chi|\right\}-\mathrm{Tr}\,\left\{\partial_{g_1} \mathcal{L} _{g_1,g}\left[\mathcal{L}_g^{-1} \right]_{\mathcal{P}_2}|\chi\rangle\langle\chi|\right\}\right|_{g_1=g}^2  \nonumber\\
&&  + 4\,  \left(\,t\,\mathrm{Tr}\,\left\{\partial^2_{g_1g_2} \mathcal{L} _{g_1,g_2} \,\mathcal{P} \,|\chi\rangle\langle\chi|\right\}-\mathrm{Tr}\,\left\{\partial^2_{g_1g_2} \mathcal{L} _{g_1,g_2}\left[\mathcal{L}_g^{-1}\right]_{\mathcal{P}_2}|\chi\rangle\langle\chi|\right\}\right)_{g_1=g_2=g}\nonumber\\
&&  + 4\, t^2 \,\mathrm{Re}\,\mathrm{Tr}\,\left\{\partial_{g_1} \mathcal{L} _{g_1,g} \,\mathcal{P} \,\partial_{g_2} \mathcal{L} _{g,g_2}\,\mathcal{P} \,|\chi\rangle\langle\chi| \right\}_{g_1=g_2=g} \nonumber\\&&-8 \,\mathrm{Re}\, \mathrm{Tr}\,\left\{\partial_{g_1} \mathcal{L} _{g_1,g} \,\mathcal{P}\,\partial_{g_2} \mathcal{L} _{g,g_2}\left[\frac{\mathcal{I}+t\mathcal{L}_g}{\mathcal{L}_g^2} \right]_{\mathcal{P}_2}|\chi\rangle\langle\chi|\right\}_{g_1=g_2=g} \nonumber\\
&&  - 8 \,\mathrm{Re} \,\mathrm{Tr}\,\left\{\partial_{g_1} \mathcal{L} _{g_1,g}\left[\frac{\mathcal{I}+t\mathcal{L}_g}{\mathcal{L}_g^2} \right]_{\mathcal{P}_2}\partial_{g_2} \mathcal{L} _{g,g_2}\, \mathcal{P}\, |\chi\rangle\langle\chi|\right\}_{g_1=g_2=g}  \nonumber\\
&& + 8 \,\mathrm{Re} \,\mathrm{Tr}\,\left\{\partial_{g_1} \mathcal{L} _{g_1,g} \left[\mathcal{L}_g^{-1} \right]_{\mathcal{P}_2}\partial_{g_2} \mathcal{L} _{g,g_2}\left[\mathcal{L}_g^{-1} \right]_{\mathcal{P}_2}|\chi\rangle\langle\chi|\right\}_{g_1=g_2=g}\nonumber\\
&& 
\quad \quad	\,+\, t^2 \, \mathcal{O} (\lambda_2 t)\,\mathcal{O}\left(c_2 (c_2+1)\, C_1^2\right) \nonumber\\
&&  \quad\quad \,+\, t\, \left\{\mathcal{O} (\lambda_2 t)\, \left[\mathcal{O}\left(c_2 \,C_1^2\,C_2\right)\,+\,\mathcal{O}\left(c_2 \,C_3\right) \right] \,+\,\mathcal{O}\left((1+c_2) \,C_1^2\,C_2\left\lVert \left[e^{t \mathcal{L}_g}\right]_{\mathcal{P}_2}\right\rVert_1\right)\,+\,\mathcal{O}\left(c_2C_1^2 C_2 \right)\mathcal{O}\left(\frac{\lambda_2}{\lambda_3}\right)\right\}\nonumber\\
\nonumber\\
&& \quad \quad\,+\,\mathcal{O}(\lambda_2 t)\,\mathcal{O}(c_2\,C_1^2\,C_2^2)+\mathcal{O}\left((1+c_2)\,C_1^2\,C_2^2\left\lVert \left[e^{t \mathcal{L}_g}\right]_{\mathcal{P}_2}\right\rVert_1\right)\,+\,\mathcal{O}\left(C_2\,C_3\left\lVert \left[e^{t \mathcal{L}_g}\right]_{\mathcal{P}_2}\right\rVert_1\right)\nonumber\\
&& \quad\quad\,+ \mathcal{O}(c_2\,C_1^2\,C_2^2)\,\mathcal{O}\left(\frac{\lambda_2}{\lambda_3}\right)\,+\,\mathcal{O}\left(C_1 \left\lVert \sum_{j\neq k, j,k>2}^{d^2} \frac{e^{t \lambda_j} -e^{t \lambda_k}}{\lambda_k(\lambda_j-\lambda_k) } |\rho_k\rangle\langle l_k |\,\partial_{g_2} \mathcal{L} _{g,g_2}\,  |\rho_j\rangle \langle l_j|  \right\rVert_1 \right)_{g_2=g}\label{sm:QFIquadratic},
\end{eqnarray}
where corrections in the approximation are given by  $c_2=\left\lVert \,|\rho_2\rangle\langle l_2|\,\right\rVert_1$, $C_1= \left\lVert \partial_{g_1}|_{g_1=g} \mathcal{L} _{g_1,g} \right\rVert_1$, $C_2 =\left\lVert \left[\mathcal{L}_g^{-1} \right]_{\mathcal{P}_2}\right\rVert_1$ and $C_3=\left\lVert \partial^2_{g_1,g_2}|_{g_1=g_2=g} \mathcal{L} _{g_1,g_2} \right\rVert_1$. We note that the estimate of the approximation error in Eq.\ \eqref{sm:QFIquadratic} is very rough and implies strong conditions on the Master dynamics to be near a DPT, i.e. the corrections to be negligible. For a particular model one should check the approximation by comparing to the exact results given by Eq.\ \eqref{sm:QFItime2}.  The quadratic terms in Eq. \eqref{sm:QFIquadratic} correspond to Eq. (7) in the paper. 

Let us note that using Eq.\ \eqref{sm:QFItimeZ} does not provide clear results in the quadratic regime. From comparing Eq.\ \eqref{sm:QFItimeAlinearZ}  to Eq.\ \eqref{sm:QFItimeAlinear}, we see that when $(-\mathrm{Re}\lambda_2)\ll (-\mathrm{Re}\lambda_3)$, many terms in Eq.\ \eqref{sm:QFItimeZ} can diverge. In order to obtain the way they diverge one needs to go back to the operators $\partial_{g_1} \mathcal{L} _{g_1,g}$ and $\partial^2_{g_1g_2} \mathcal{L} _{g_1,g_2}$ and therefore to Eqs.\ \eqref{sm:QFItime2} and \eqref{sm:QFIquadratic}. \\

\emph{Quadratic behaviour at a first-order DPT}. At a first-order DPT we have $\lambda_2=0$ and the considered time regime is infinitely long due to $\tau=\infty$. Since we have $\lim_{t\rightarrow\infty}\left\lVert \left[e^{t\mathcal{L}_g}\right]_{\mathcal{P}_2}\right\rVert_1=0$, in the limit of long time $t$ Eq.\ \eqref{sm:QFIquadratic} becomes exact, as all the corrections disappear. Therefore, Eq.\ \eqref{sm:QFIquadratic}  gives asymptotic quadratic behaviour of the QFI: 
\begin{eqnarray}
\lim_{t\rightarrow\infty} t^{-2}F(|\Psi_g(t)\rangle )= -4\left|\mathrm{Tr}\,\left\{\partial_{g_1} \mathcal{L} _{g_1,g}\,\mathcal{P} \,|\chi\rangle\langle\chi|\right\}\right|_{g_1=g}^2 + 4 \,\mathrm{Re}\,\mathrm{Tr}\,\left\{\partial_{g_1} \mathcal{L} _{g_1,g} \,\mathcal{P} \,\partial_{g_2} \mathcal{L} _{g,g_2}\,\mathcal{P} \,|\chi\rangle\langle\chi| \right\}_{g_1=g_2=g},\label{sm:QFIatDPTquadratic},
\end{eqnarray}
see also Eq. (7) in the paper. \\

\subsection{D. \, Quadratic behaviour and bimodality} 
Here we consider a system at a first-order DPT with respect to photon counting statistics. We show how the quadratic behaviour of the QFI emerges from the bimodality of the total photon number which which the generator of the phase transformation. This relation is given by Eq. (10) in the paper.  A similar behaviour holds in the general case of arbitrary parameter defence, but this will be discussed in later work~\cite{NextPaper}.

We consider the estimated parameter to be the phase $g=\phi$ encoded on photons emitted by a system, which can be formalised by defining $L_{1,\phi}=e^{-i\phi}L_1$, where $L_1$ is a jump operator of the Master operator $\mathcal{L}$, see the paper. In that case the Master operator  does not depend on  $\phi$, $\mathcal{L}_\phi=\mathcal{L}$. We consider the system to be at a first-order DPT, and for simplicity we restrict to the case where the zero eigenvalue of the Master operator $\mathcal{L}$ has degeneracy two. This means that there exist two stationary states $\rho_{\rm A}$ and $\rho_{\rm I}$ which are supported on orthogonal subspaces $\mathcal{H}_{\rm A}$, $\mathcal{H}_{\rm I}$, so that  $\mathcal{H}=\mathcal{H}_{\rm A}\oplus\mathcal{H}_{\rm I}$. Moreover, the jump and hamiltonian operators have a block diagonal form in this decomposition 
$H=H^{\rm A}\oplus H^{\rm I}$, $L_{1,\phi}=L^{\rm A}_{1,\phi}\oplus L^{\rm I}_{1,\phi}$ and $L_{j}=L^{\rm A}_{j}\oplus L^{\rm I}_{j}$, $j=2,...,k$, acting on $\mathcal{H}=\mathcal{H}_{\rm A}\oplus\mathcal{H}_{\rm I}$. Let $\mathcal{P}_{\mathcal{H}_{\rm A}}$, $\mathcal{P}_{\mathcal{H}_{\rm I}}$ denote orthogonal projections on $\mathcal{H}_{\rm A}$, $\mathcal{H}_{\rm I}$, respectively. 

 We note that  the block-diagonal structure is preserved for $\partial_{\phi_2}|_{\phi_2=\phi}\mathcal{L} _{\phi,\phi_2} \rho = i L_1 \rho L_1^\dagger$, see Eq. (8) in the paper. Thus, $\mathrm{Tr}\left\{ \mathcal{P}_{\mathcal{H}_{\rm I}}\, L_1 \rho_{\rm A} L_1^\dagger\right\}=0=\mathrm{Tr}\left\{ \mathcal{P}_{\mathcal{H}_{\rm A}}\, L_1 \rho_{\rm I} L_1^\dagger\right\}$, which leads to the following simple form of the asymptotic QFI:
\begin{eqnarray}
\lim_{t\rightarrow\infty} t^{-2}F(|\Psi_\phi(t)\rangle )&=& -4\left|\mathrm{Tr}\,\left\{\partial_{\phi_1} \mathcal{L} _{\phi_1,\phi}\,\mathcal{P} \,|\chi\rangle\langle\chi|\right\}\right|_{\phi_1=\phi}^2 + 4 \,\mathrm{Re}\,\mathrm{Tr}\,\left\{\partial_{\phi_1} \mathcal{L} _{\phi_1,\phi} \,\mathcal{P} \,\partial_{\phi_2} \mathcal{L} _{\phi,\phi_2}\,\mathcal{P} \,|\chi\rangle\langle\chi| \right\}_{\phi_1=\phi_2=\phi}\nonumber\\
&=&-4\left(p_{\rm A} \mathrm{Tr}\,\left\{  L_1^\dagger L_1 \rho_{\rm A}\right\}+ p_{\rm I}\mathrm{Tr}\,\left\{ L_1^\dagger L_1 \rho_{\rm I} \right\}\right)^2 + 4 p_{\rm A}  \left( \mathrm{Tr}\,\left\{  L_1^\dagger L_1 \rho_{\rm A}\right\}\right)^2\,+\,4 p_{\rm I} \left( \mathrm{Tr}\,\left\{  L_1^\dagger L_1 \rho_{\rm I}\right\}\right)^2
\nonumber\\
&=&  4\, p_{\rm A} p_{\rm I}\left(\mathrm{Tr}\,\left\{  L_1^\dagger L_1 \rho_{\rm A}\right\}-\mathrm{Tr}\,\left\{ L_1^\dagger L_1 \rho_{\rm I} \right\}\right)^2\label{sm:QFIt2},
\end{eqnarray}
where $p_{\rm A}=\mathrm{Tr}\,\left\{\mathcal{P}_{\mathcal{H}_{\rm A}}\,|\chi\rangle\langle\chi|\right\}$ and $p_{\rm I}=\mathrm{Tr}\,\left\{\mathcal{P}_{\mathcal{H}_{\rm I}}\,|\chi\rangle\langle\chi|\right\}$.

We will now show how the stationary states $\rho_{\rm A}$, $\rho_{\rm I}$ correspond to the bimodal distribution of the generator, which is the total photon count $\Lambda(t)$, when measured on $|\Psi(t)\rangle$. For $|\Psi_\phi(t)\rangle=e^{-i\phi\Lambda(t)}|\Psi(t)\rangle$, we have $\langle \Psi_{\phi_{1}}(t)| \Psi_{\phi_{2}}(t)\rangle
=\langle \Psi(t)| e^{  i\,(\phi_{1}-\phi_{2}) \Lambda(t)}\,| \Psi(t)\rangle$. We can then express the photon emission average as follows:
\begin{eqnarray*}
\langle\Psi(t)| \Lambda(t)|\Psi(t)\rangle&=&  i\partial_{\phi_2} \langle \Psi_{\phi}(t)| \Psi_{\phi_{2}}(t)\rangle_{\phi_2=\phi}= i\,\partial_{\phi_2} \mathrm{Tr}\{e^{t\mathcal{L}_{\phi,\phi_2}}|\chi\rangle\langle\chi |\}_{\phi_2=\phi},\quad\mathrm{ and \,thus}\\\\
\lim_{t\rightarrow\infty} t^{-1}\langle\Psi(t)| \Lambda(t)|\Psi(t)\rangle &=&  i\,\mathrm{Tr}\,\left\{\partial_{\phi_2} \mathcal{L} _{\phi,\phi_2}\,\mathcal{P} \,|\chi\rangle\langle\chi | \right\}_{\phi_2=\phi} = - \mathrm{Tr}\,\left\{ L_1^\dagger L_1\mathcal{P} \,|\chi\rangle\langle\chi | \right\} .
\end{eqnarray*}
We represent the initial state of the system as $|\chi\rangle=\sqrt{p}|\chi_{\rm A}\rangle+\sqrt{1-p}|\chi_{\rm I}\rangle$, where  $|\chi_{\rm A}\rangle\in\mathcal{H}_{\rm A}$ and $|\chi_{\rm I}\rangle\in\mathcal{H}_{\rm A}$ and $0\leq p\leq1$. By setting $p=1$ or $p=0$ we arrive at $-\mathrm{Tr}\,\left\{ L_1^\dagger L_1\, \rho_{\rm A}\right\}=\mu_{\rm A}$ and  $-\mathrm{Tr}\,\left\{ L_1^\dagger L_1\, \rho_{\rm I}\right\}=\mu_{\rm I}$, where  $\mu_{\rm A}$, $\mu_{\rm I}$ are the asymptotic rates of $\Lambda(t)$ when the system is initially in the state $|\chi_{\rm A}\rangle\in\mathcal{H}_{\rm A}$, $|\chi_{\rm I}\rangle\in\mathcal{H}_{\rm I}$, respectively. Therefore Eq.\ \eqref{sm:QFIt2} becomes:
\begin{equation}
\lim_{t\rightarrow\infty} t^{-2}F( |\Psi_{g}(t)\rangle ) = 4\, p_{\rm A} p_{\rm I} \left( \mu_{\rm A} - \mu_{\rm I} \right)^{2},\label{sm:QFIt2Phases}
\end{equation}
which corresponds to Eq. (10) in the paper. Let us assume $\mu_{\rm A}>\mu_{\rm I}$. We see that in the asymptotic limit $t\gg \tau'$ we can define two \emph{dynamical phases} corresponding to active (A) and inactive (I) mode in total photon count $\Lambda(t)$ distribution, to be any MPS states $|\Psi^{\rm A}(t)\rangle$, $|\Psi^{\rm I}(t)\rangle$, which after tracing out the output are supported only on $\mathcal{H}_{\rm A}$, $\mathcal{H}_{\rm I}$, respectively. In order to ensure quadratic scaling of the QFI $F(|\Psi_\phi(t)\rangle)$, the initial state $|\chi\rangle$ of the system needs to be a superposition of states from $\mathcal{H}_{\rm A}$ and $\mathcal{H}_{\rm I}$, so that both $p_{\rm A},p_{\rm I}>0$. When the asymptotic emission rates are equal $\mu_{\rm A}=\mu_{\rm I}$, there is no quadratic scaling of the QFI, as  the asymptotic distribution of photon counts $\Lambda(t)$ is unimodal with variance scaling linearly with time $t$. \\

\emph{Quadratic behaviour and approximate bimodality for an intermittent system}. Near a DPT in photon emissions, the system dynamics is intermittent and switches between long time intervals with different emission rates. The typical length of those intervals is given by the correlation time $\tau=(-\mathrm{Re} \lambda_2)^{-1}$. Therefore, we expect to be able to construct approximate stationary states for the quadratic QFI regime $\tau'\ll t\ll\tau$. Let us note that  the Master operator $\mathcal{L}$ has only one stationary state $\rho_{ss}$ of $\mathcal{L}$ and its second eigenvector $\rho_2$ fulfills $\mathrm{Tr}\rho_2=0$ due to normalisation of eigenvectors.  Nevertheless, $\mathcal{L}$ is degenerate up to order $\lambda_2$. Below we briefly present a construction of two approximately stationary states with different emission rates as linear combinations of $\rho_{ss}$ and $\rho_2$. The construction closely follows the  theory of classical non-equilibrium first-order phase transitions~\cite{Gaveau1998}. We leave rigorous proofs and discussion of Eq.\ \eqref{sm:QFIt2Phases} in that case for later work~\cite{NextPaper}. 

Consider first the case when  a first-order DPT is approached by changing parameters in the Master operator $\mathcal{L}$, which will then guide the more general considerations near a DPT. When approaching the DPT, the first two eigenvectors converge to $\rho_{1}$ and $\rho_2$, such that $\rho_1\geq 0$, $\mathrm{Tr}\rho_{1}=1$ and $\mathrm{Tr}\rho_{2}=0$. One can show that in that case $\rho_{1}=p\,\rho_{\rm A}+(1-p)\,\rho_{\rm I}$, $\rho_{2}=\rho_{\rm A}-\rho_{\rm I}$ and $l_2=(1-p)\,\mathcal{P}_{\mathcal{H}_{\rm A}}-p\,\mathcal{P}_{\mathcal{H}_{\rm I}}$, where $0<p<1$, $\rho_{\rm A}$, $\rho_{\rm I}$ are the stationary states supported on orthogonal subspaces $\mathcal{H}_{\rm A}$, $\mathcal{H}_{\rm I}$, respectively, and $\mathcal{P}_{\mathcal{H}_{\rm A}}$, $\mathcal{P}_{\mathcal{H}_{\rm I}}$ are the orthogonal projections on these subspaces.  In the case of the system near a DPT, the construction of approximate stationary states is as follows. The Master operator $\mathcal{L}$ acts almost block-diagonally, i.e. $H$, $L_j$, $j=1,...,k$ are approximately block-diagonal with respect to an orthogonal splitting into subspaces $\mathcal{H}_1$ and $\mathcal{H}_2$. For pure initial states $|\chi^{(1)}\rangle$, $|\chi^{(2)}\rangle$ supported in these subspaces, the corresponding evolved states $\rho^{(1)}(s)$, $\rho^{(2)}(s)$ will still be supported in the blocks for time $s$ in quadratic regime of QFI, $\tau'\ll s\ll \tau$,  and will be well approximated by linear combinations of $\rho_{ss}$ and $\rho_2$ (due to $s\gg\tau' $).

 Moreover, $\rho^{(1)}(s)$, $\rho^{(2)}(s)$ are almost stationary w.r.t $e^{t\mathcal{L}}$, where $t=\mathcal{O}(s)$. In order to define the approximate blocks, $\mathcal{H}_1$, $\mathcal{H}_2$, and the initial states $|\chi^{(1)}\rangle$, $|\chi^{(2)}\rangle$ using the Master operator $\mathcal{L}$, we assume $\lambda_2\in\mathbb{R}$  for simplicity. In this case both $\rho_2$ and $l_2$ are Hermitian matrices on $\mathcal{H}$, i.e. they diagonalise and their spectra are real. First, inspired by the form of the second eigenvector at a first-order DPT, $l_2=(1-p)\,\mathcal{P}_{\mathcal{H}_{\rm A}}-p\,\mathcal{P}_{\mathcal{H}_{\rm I}}$, we define the subspaces $\mathcal{H}_1$, $\mathcal{H}_2$ in the following way. $\mathcal{H}_1$ is spanned by the eigenvectors of $l_2$ which correspond to positive eigenvalues close to the maximal eigenvalue of $l_2$, while $\mathcal{H}_2$ is spanned by eigenvectors of $l_2$ corresponding to negative eigenvalues close to the minimal eigenvalue of $l_2$. Next, the initial states $|\chi^{(1)}\rangle$ and $|\chi^{(2)}\rangle$ are chosen to be the eigenvectors corresponding to maximal and minimal eigenvalue of $l_2$, respectively. Finally, the two approximate dynamical phases in photon emissions are defined as any MPS states which after tracing out the output are supported mostly on $\mathcal{H}_1$, $\mathcal{H}_2$, respectively. Using the above decomposition, it can be shown that the quadratic behaviour of the QFI is again related to the two approximate modes in the counting distribution given by the approximate dynamical phases, see Eq. (10) in the paper.  \\

\end{document}